\documentclass[preprint,superscriptaddress,showpacs,preprintnumbers,amsmath,amssymb]{revtex4}

\usepackage{graphicx,color}
\usepackage{epsfig}
\usepackage{dcolumn}
\usepackage{bm}
\usepackage{CJK}

\begin{document}
\begin{CJK*}{GBK}{song}

\title{Collective Hamiltonian for chiral modes}

\author{Q. B. Chen}
\affiliation{State Key Laboratory of Nuclear Physics and Technology, School of Physics,
             Peking University, Beijing 100871, China}%

\author{S. Q. Zhang}\email{sqzhang@pku.edu.cn}
\affiliation{State Key Laboratory of Nuclear Physics and Technology, School of Physics,
             Peking University, Beijing 100871, China}%

\author{P. W. Zhao}
\affiliation{State Key Laboratory of Nuclear Physics and Technology, School of Physics,
             Peking University, Beijing 100871, China}%

\author{R. V. Jolos}
\affiliation{Joint Institute for Nuclear Research, Dubna 141980 , Russia}%
\affiliation{Institut f\"{u}r Kernphysik der Universit\"{a}t zu K\"{o}ln, K\"{o}ln 50937, Germany}%

\author{J. Meng}\email{mengj@pku.edu.cn}
\affiliation{State Key Laboratory of Nuclear Physics and Technology, School of Physics,
             Peking University, Beijing 100871, China}%
\affiliation{School of Physics and Nuclear Energy Engineering, Beihang University, Beijing 100191, China}%
\affiliation{Department of Physics, University of Stellenbosch, Stellenbosch, South Africa}%

\date{\today}

\begin{abstract}
A collective model is proposed to describe the chiral rotation and
vibration and applied to a system with one $h_{11/2}$ proton
particle and one $h_{11/2}$ neutron hole coupled to a triaxial rigid
rotor. The collective Hamiltonian is constructed from the potential
energy and mass parameter obtained in the tilted axis cranking
approach. By diagonalizing the collective Hamiltonian with a box
boundary condition, it is found that for the chiral rotation, the
partner states become more degenerate with the increase of the
cranking frequency, and for the chiral vibrations, their important
roles for the collective excitation are revealed at the beginning of
the chiral rotation region.
\end{abstract}
\date{\today}

\maketitle


\section{Introduction}\label{sec1}

Since the chirality in nuclear physics was originally suggested by
Frauendorf and Meng in 1997~\cite{Frauendorf1997NPA}, it has been
one of the hot topics in nuclear physics. The corresponding
experimental signals, the chiral doublet bands,  were first
observed in $N = 75$ isotones in 2001~\cite{Starosta2001PRL}, and so
far have been extensively reported in more than 30 nuclei.

For a rotating nucleus with triaxiality, the three angular momenta,
including collective rotor angular momentum along the intermediate
axis, together with angular momenta of the valence particles (holes)
along the nuclear short (long) axis, can construct the chiral
geometry~\cite{Frauendorf1997NPA, Starosta2001PRL}. It has been
demonstrated that the chiral doublet bands generally start from a
left-right chiral vibration mode and evolve into a static chiral
mode with the increase of spin~\cite{Mukhopadhyay2007PRL,
B.Qi2009PLB}.

Theoretically, chiral doublet bands were first predicted by the
tilted axis cranking (TAC) approach and  particle rotor model
(PRM)~\cite{Frauendorf1997NPA}. Subsequently, numerous efforts have
been devoted to the development of both the TAC
methods~\cite{Dimitrov2000PRL, Olbratowski2004PRL,
Olbratowski2006PRC} and PRM models~\cite{J.Peng2003PRC,
S.Q.Zhang2007PRC, B.Qi2009PLB}. Both PRM and TAC have their own
advantages and disadvantages. The PRM is a quantal model consisting
of the collective rotation and the intrinsic single-particle
motions, and the total angular momentum is a good quantum number. It
describes a system in the laboratory reference frame where the
spontaneously broken chiral symmetry in the intrinsic reference
frame is restored. In the PRM, the energy splitting and quantum
tunneling between the doublet bands can be obtained directly.
However,  rigid rotor with quadrupole deformation parameters $\beta$
and $\gamma$  has to be assumed at the very beginning in the
practical PRM calculation.

TAC approach is based on mean-field theory, and permits the
calculation for the orientation of the density distribution relative
to the angular momentum vector. However the TAC cannot describe the
energy splitting and the quantum tunneling between the chiral
doublet bands~\cite{Frauendorf1997NPA}. Up to now, the TAC method
based on the Woods-Saxon or Nilsson
potential~\cite{Dimitrov2000PRL}, as well as the microscopic
self-consistent Skyrme Hartree-Fock model~\cite{Olbratowski2004PRL,
Olbratowski2006PRC} have been devoted to study the chirality.

To describe the energy splitting between the chiral doublet bands,
one has to go beyond the mean-field approximation. So far, this has
been done in the framework of tilted axis cranking plus the random
phase approach (RPA) model~\cite{Mukhopadhyay2007PRL,
Almehed2011PRC}. However, this model is restricted in the
description of the chiral vibration since the quantum tunneling in
the chiral rotation is beyond the realm of RPA. Therefore, it is
imperative to search a unified method for studying both chiral
rotation and vibration in the framework of TAC.

In analogy to Bohr Hamiltonian, constructing a collective
Hamiltonian on the TAC solutions provides one of the ways out.
Instead of the $\beta$ and $\gamma$ degrees of freedom in Bohr
Hamiltonian, a chiral degree of freedom should be introduced.
Subsequently, the quantal tunneling in the region of chiral rotation
can be described by considering the chiral fluctuations around
mean-field minima besides the region of chiral vibration naturally
described by the collective Hamiltonian.

In this work the collective Hamiltonian for a system of one
$h_{11/2}$ proton particle and one $h_{11/2}$ neutron hole coupled
to a triaxial rigid rotor is constructed. The potential energy and
mass parameters involved in the collective Hamiltonian are extracted
from the TAC calculations. By diagonalizing the collective
Hamiltonian, the energy levels and wave functions are calculated and
discussed in detail. The paper is organized as follows. In
Sec. \ref{sec2}, based on the tilted axis cranking approach, the
procedures for constructing and solving the collective Hamiltonian
are introduced. The numerical details are presented in Sec.
\ref{sec3}. In Sec. \ref{sec4}, the potential energy and mass
parameters obtained from TAC approach are shown and the
corresponding energy levels and wave functions obtained from the
collective Hamiltonian are discussed in details. Finally, the
summary is given in Sec. \ref{sec5}.


\section{Theoretical framework}\label{sec2}

\subsection{Tilted axis cranking model}

The detailed formalism for TAC can be found in
Ref.~\cite{Frauendorf1993NPA}. For a schematic discussion, similar
as Ref.~\cite{Frauendorf1997NPA}, we consider a system of the
$h_{11/2}$ proton particle and the $h_{11/2}$ neutron hole coupled
to a triaxial rigid rotor. The cranking Hamiltonian reads
\begin{eqnarray}\label{eq1}
  \hat{h}^\prime &=& \hat{h}_{\rm def}-\vec{\omega}\cdot\hat{\vec{j}},\\
  \hat{\vec{j}}  &=& \hat{\vec{j}}_\pi+\hat{\vec{j}}_\nu,\\
  \vec{\omega}   &=& (\omega\sin\theta\cos\varphi, \omega\sin\theta\sin\varphi, \omega\cos\theta),
\end{eqnarray}
where the Hamiltonian of the deformed field is $\hat{h}_{\rm
def}=\hat{h}_{\rm def}^{\rm \pi}+\hat{h}_{\rm def}^{\rm \nu}$ with
the single-$j$ shell Hamiltonian
\begin{equation}\label{eq2}
\hat{h}_{\rm def}^{\rm\pi(\nu)} = \pm\frac{1}{2}C\Big\{(\hat{j}_3^2-\frac{j(j+1)}{3})\cos\gamma
                                + \frac{1}{2\sqrt{3}}(\hat{j}_+^2+\hat{j}_-^2)\sin\gamma\Big\}.
\end{equation}

The TAC solutions are obtained self-consistently by minimizing the total Routhian surface
\begin{equation}\label{eq3}
  E^\prime(\theta,\varphi)=\langle h^\prime\rangle-\frac{1}{2}\sum_{k=1}^3\mathcal{J}_k \omega_k^2,
\end{equation}
with respect to the angles $\theta$ and $\varphi$, where the moments
of inertia for irrotational flow are adopted, i.e.,
\begin{equation}
  \mathcal{J}_k = \mathcal{J}_0\sin^2(\gamma-\displaystyle\frac{2\pi}{3}k).
\end{equation}

As discussed in Ref.~\cite{Frauendorf1997NPA}, there are three kinds
of different solutions which can be distinguished by the different
values of $\theta$ and $\varphi$: a) Principal axis cranking (PAC)
solution: $\theta=0, \pi/2$, $\varphi=0, \pm\pi/2$. b) Planar TAC
solution: $\theta\neq 0,\pm\pi/2$, $\varphi=0, \pm\pi/2$ or
$\theta=\pi/2$, $\varphi\neq 0, \pm\pi/2$.  c) Aplanar TAC solution:
$\theta\neq 0, \pi/2$, $\varphi\neq 0, \pm\pi/2$, i.e., the chiral
solution.

In Ref.~\cite{Starosta2001NPA}, an orientation operator
$\hat{\sigma}$ was introduced to characterize the chiral degree of
freedom. In the TAC approach for chiral solutions, the orientation
operator is written as
\begin{equation}
  \hat{\sigma} = (\hat{\vec{j}}_\pi\times \hat{\vec{j}}_\nu)\cdot \hat{\vec{R}}
               = |\hat{\vec{j}}_\pi|\cdot|\hat{\vec{j}}_\nu|\cdot |\hat{\vec{R}}|\cdot \sin\theta_{\rm PN}
                 \cdot \sin\theta\sin\varphi,
\end{equation}
in which $\theta_{\rm PN}$ denotes the angle between of
$\hat{\vec{j}}_\pi$ and $\hat{\vec{j}}_\nu$. This operator has
opposite signs for the left-handed and the right-handed systems.
Follow Ref.~\cite{Frauendorf1997NPA}, we restrict the angle $\theta$
to $0\leq \theta \leq \pi/2$, and by varying $\varphi$ from $-\pi/2$
to $\pi/2$, we get $\varphi$ as a dynamical variable describing a
transition from the left-handed to the right-handed system. Thus for
a given $\theta$, the angle $\varphi$ can be used to characterize
the chiral degree of freedom. We note that the calculations
performed in Refs.~\cite{Dimitrov2000PRL} and \cite{Almehed2011PRC}
for the potential energy as a function of $\theta$ and $\varphi$ for
$^{134}\rm Pr$ have shown a softer nature of the potential in the
$\varphi$ direction compare to the $\theta$ direction.

\subsection{Collective Hamiltonian}

Based on the TAC model, a collective Hamiltonian including the
chiral degree of freedom could be constructed. The classical form of
a collective Hamiltonian can be written in terms of $\varphi$ as
\begin{equation}\label{eq11}
H_{\rm coll} = T_{\rm vib}(\varphi)+V(\varphi),
\end{equation}
where $\varphi$ is the variable which characterizes the chirality,
and the potential energy $V(\varphi)$ could be extracted by
minimizing the total Routhian Eq. (\ref{eq3}) with respect to
$\theta$ for given $\varphi$.

The vibrational kinetic energy of the chiral degree of freedom can be written as
\begin{equation}\label{eq15}
T_{\rm vib} = \frac{1}{2}B\dot{\varphi}^2,
\end{equation}
where $B$ is the mass parameter. The Hamiltonian (\ref{eq11}) is
quantized according to the general Pauli prescription
\cite{Pauli1993HdP}, i.e., for a classical kinetic energy,
\begin{equation}
T = \frac{1}{2}\sum_{ij}B_{ij}(q)\dot{q}_i\dot{q}_j,
\end{equation}
the corresponding quantized form reads
\begin{equation}
  \hat{H}_{\rm kin} = -\frac{\hbar^2}{2}\frac{1}{\sqrt{\det{B}}}\sum_{ij}\frac{\partial}{\partial q_i}
                      \sqrt{\det{B}}(B^{-1})_{ij}\frac{\partial}{\partial q_j}.
\end{equation}
For the vibrational kinetic energy in Eq. (\ref{eq15}), the mass
parameter takes a $1\times 1$ matrix form. Thus, the corresponding
quantized form reads
\begin{equation}
  \hat{H}_{\rm kin} = -\frac{\hbar^2}{2\sqrt{B(\varphi)}}\frac{\partial}{\partial\varphi}\frac{1}
                      {\sqrt{B(\varphi)}}\frac{\partial}{\partial \varphi}.
\end{equation}
Therefore, the quantized form of the collective Hamiltonian in Eq. (\ref{eq11}) turns out to be
\begin{equation}\label{eq13}
  \hat{H}_{\rm coll} = -\frac{\hbar^2}{2\sqrt{B(\varphi)}}\frac{\partial}{\partial\varphi}
                        \frac{1}{\sqrt{B(\varphi)}}\frac{\partial}{\partial \varphi}+V(\varphi).
\end{equation}
Note that the volume element in the present collective space is
\begin{equation}\label{eq14}
  \int d\tau_{\rm coll} = \int d\varphi \sqrt{B(\varphi)},
\end{equation}
and the quantized Hamiltonian of Eq.~(\ref{eq13}) is Hermitian with
respect to the collective measure in Eq. (\ref{eq14}).

\subsection{Mass parameter}

The details for calculating the mass parameter can be found
in~\cite{Rowe1970Book}. In this subsection, for completeness, a
brief procedure is presented. Considering a variable $\varphi$ as a
function of time, the time dependent Schr\"{o}dinger equation is
written as
\begin{equation}\label{eq4}
  \hat{h}^\prime(t)|\psi(t)\rangle = i\hbar\frac{\partial}{\partial t}|\psi(t)\rangle.
\end{equation}
The general solution of this equation is
\begin{equation}\label{eq5}
  |\psi(t)\rangle = \sum_k a_k(t)e^{i\phi_k(t)}|k\rangle,
\end{equation}
where $\phi_k(t) = -\displaystyle \frac{1}{\hbar}\int_0^t
E_k(t^\prime)\textrm{d}t^\prime$ and $E_k(t)$ is the eigenvalue of
the Hamiltonian $\hat{h}^\prime(t)$. By substituting Eq. (\ref{eq5})
into Eq. (\ref{eq4}), we obtain
\begin{equation}\label{eq6}
  \dot{a}_l = -\dot{\varphi}\sum_k a_k(t)e^{i(\phi_k-\phi_l)}\langle l|\frac{\partial}{\partial \varphi}|k\rangle.
\end{equation}

We assume that the wave function is the lowest eigenstate of the
cranking Hamiltonian $\hat{h}^\prime$ in Eq.~(\ref{eq1}) at $t=0$,
i.e., $a_k=\delta_{k0}$ at $t=0$. Here, $|0\rangle$ means the lowest
state of $\hat{h}^\prime$. If we further assume that $a_{l\neq 0}$
are small and the variable $\varphi$ follows the relation of
$\ddot{\varphi}=-\Omega^2\varphi$ with $\Omega$ being the
vibrational frequency, the solution of the Eq. (\ref{eq6}) can be
obtained as
\begin{equation}\label{eq7}
  a_l = \frac{i\hbar(E_l-E_0)\dot{\varphi}+\hbar^2\Omega^2\varphi}{(E_l-E_0)^2-\hbar^2\Omega^2}
        \langle l|\frac{\partial}{\partial \varphi}|0\rangle e^{-i(\phi_l-\phi_0)},
\end{equation}
where $E_l$ and $E_0$ are the eigen energies of the cranking Hamiltonian $\hat{h}^\prime$ in Eq.~(\ref{eq1}).

The energy of the system can be written as
\begin{eqnarray}\label{eq8}
  E(t) &=& E_0+\sum_{l\neq 0}(E_l-E_0)|a_l|^2\nonumber\\
       &=& E_0+\sum_{l\neq 0}(E_l-E_0)\frac{\Big[\hbar^2(E_l-E_0)^2\dot{\varphi}^2+(\hbar\Omega)^4\varphi^2\Big]
           |\langle l|\frac{\partial}{\partial \varphi}|0\rangle|^2}{[(E_l-E_0)^2-\hbar^2\Omega^2]^2}\nonumber\\
       &=& E_0+\frac{1}{2}B(\varphi)\dot{\varphi}^2.
\end{eqnarray}

Neglecting the high-order terms ($\propto\varphi^2$), the mass parameter can be obtained as
\begin{eqnarray}\label{eq9}
  B(\varphi) &=& 2\hbar^2\sum_{l\neq 0}\frac{(E_l-E_0)^3|
                  \langle l|\frac{\partial}{\partial \varphi}|0\rangle|^2}
                  {[(E_l-E_0)^2-\hbar^2\Omega^2]^2}\nonumber\\
             &=& 2\hbar^2\sum_{l\neq 0}\frac{(E_l-E_0)
                 |\langle l|[\hat{h}^\prime,\frac{\partial}{\partial\varphi}]|0\rangle|^2} {[(E_l-E_0)^2-\hbar^2\Omega^2]^2}.
\end{eqnarray}

By substituting Eq.~(\ref{eq1}) into Eq.~(\ref{eq9}) and
using the relation
\begin{equation}
  \langle l|[h_{\rm def}-\vec{\omega}\cdot\hat{\vec{j}},\frac{\partial}{\partial\varphi}]|0\rangle
  = \frac{\partial \vec{\omega}}{\partial\varphi}\langle l|\hat{\vec{j}}|0\rangle,
\end{equation}
one obtains,
\begin{equation}\label{eq16}
  B(\varphi) = 2\hbar^2\sum_{l\neq 0}\frac{(E_l-E_0)\Big| \frac{\partial\vec{\omega}}{\partial \varphi}
               \langle l|\hat{\vec{j}}|0\rangle\Big|^2}{[(E_l-E_0)^2-\hbar^2\Omega^2]^2}.
\end{equation}

Once the vibrational frequency $\Omega$ is known, the mass parameter
$B(\varphi)$ can be then determined by Eq. (\ref{eq16}). Normally,
$\Omega$ is determined as follows: a) for the chiral rotations, it
is approximately taken as zero because the barrier penetration
between the left-handed and right-handed states is low; b) for the
chiral vibrations, the potential energy $V(\varphi)$ can be
approximated by the harmonic oscillator potential
$\frac{1}{2}K\varphi^2$ with the corresponding spring coefficient
$K$. The mass parameter is
\begin{equation}\label{eq10}
  B = \frac{K}{\Omega^2}.
\end{equation}
Combining Eqs. (\ref{eq16}) and (\ref{eq10}), the value of
$\Omega$ can be obtained for the chiral vibrations.

\subsection{Basis space}

After the mass parameter $B(\varphi)$ is obtained from the TAC
model, the collective Hamiltonian (\ref{eq13}) is constructed.
It is easy to find that the collective Hamiltonian Eq. (\ref{eq13})
keeps the parity conservation with respect to $\varphi$$\to$$
-\varphi$. Therefore, the eigenstates of the collective Hamiltonian
can be divided into two separate subspaces, i.e., the positive
parity states and the negative parity states.

For the positive parity, the basis states can be taken as
\begin{equation}
  \psi_n(\varphi) = \sqrt{\displaystyle \frac{2}{\pi}}\displaystyle \frac{\cos (2n-1)\varphi}{B^{1/4}(\varphi)},
                    \quad n\geq 1
\end{equation}
and for the negative parity they are
\begin{equation}
  \psi_n(\varphi) = \sqrt{\displaystyle \frac{2}{\pi}}\displaystyle \frac{\sin 2n\varphi}{B^{1/4}(\varphi)},
                    \quad n\geq 1
\end{equation}
These basis states fulfill the box boundary condition,
\begin{equation}\label{eq17}
\psi_n(\pi/2) = \psi_n(-\pi/2) = 0.
\end{equation}
For the chosen basis, the normalization conditions with respect to
the collective measure in Eq. (\ref{eq14}) are satisfied.

The wave function can be then expanded by the basis states as
\begin{equation}
  \psi(\varphi) = \sum_{n=1}^\infty a_n\sqrt{\frac{2}{\pi}}\frac{\cos (2n-1)\varphi}{B^{1/4}(\varphi)}
                + \sum_{n=1}^\infty b_n\sqrt{\frac{2}{\pi}}\frac{\sin 2n\varphi}{B^{1/4}(\varphi)},
\end{equation}
where the expansion coefficients $a_n$ and $b_n$ ($n\geq 1$) are
obtained by diagonalizing the collective Hamiltonian Eq.~(\ref{eq13}).


\section{Numerical details}\label{sec3}

In the following calculations, the symmetric particle-hole
configuration $\pi h_{11/2}\otimes \nu h_{11/2}^{-1}$ is considered
and $\gamma$ deformation is assumed as $\gamma=-30^\circ$. The
coupling parameters $C$ in the single-$j$ shell Hamiltonian are
taken as $C_\pi=0.25~\rm MeV$ for the proton particle and
$C_\nu=-0.25~\rm MeV$ for the neutron hole, respectively. The
moment of inertia is chosen as $\mathcal{J}_0=40~\hbar^2/\rm MeV$.
These parameters are the same as in Ref.~\cite{Frauendorf1997NPA}.


\section{Results and discussion}\label{sec4}

\subsection{Collective potential}\label{se1}

\begin{figure}[!th]
\begin{center}
  \includegraphics[width=7.5 cm]{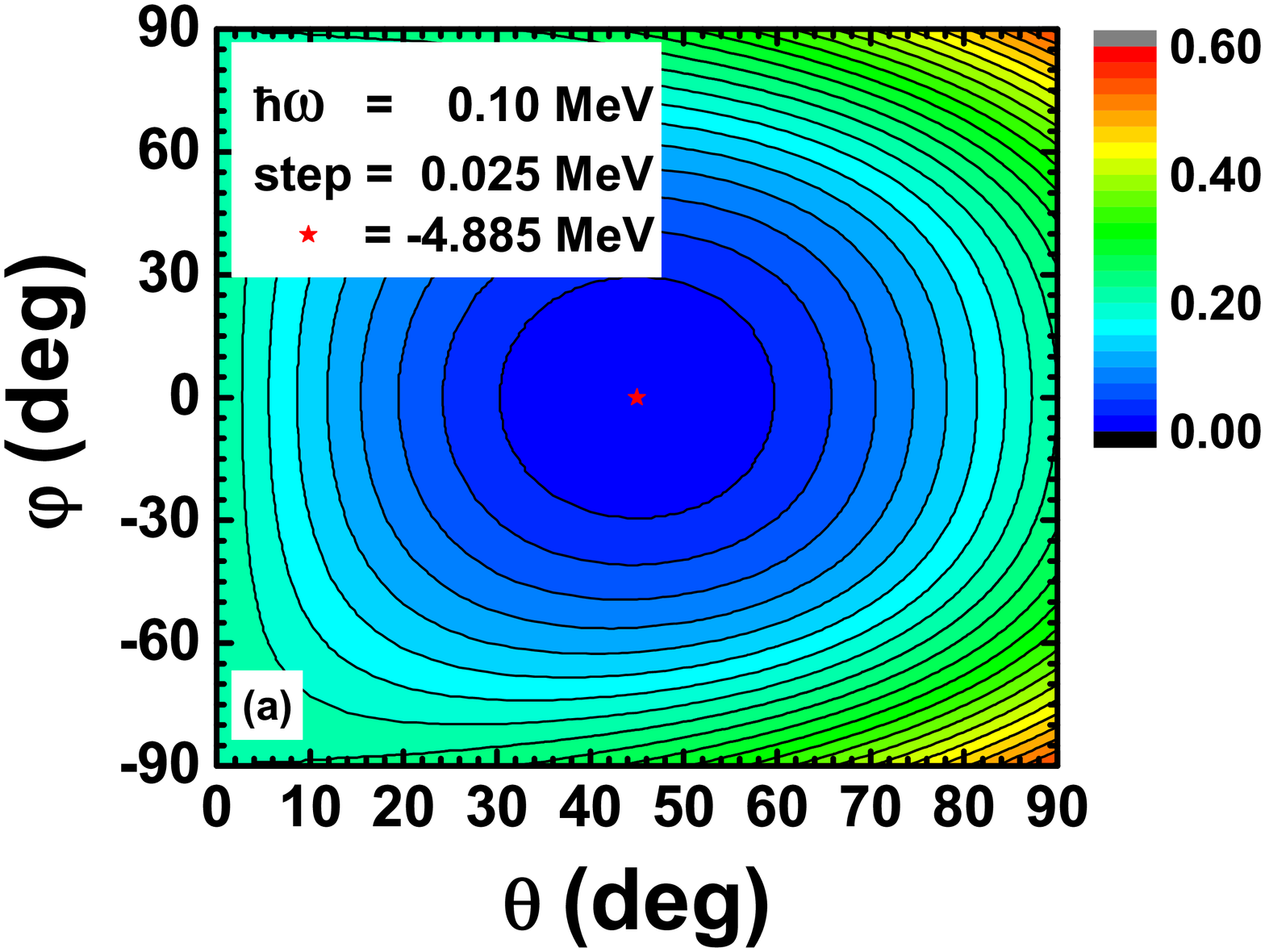}
  \includegraphics[width=7.5 cm]{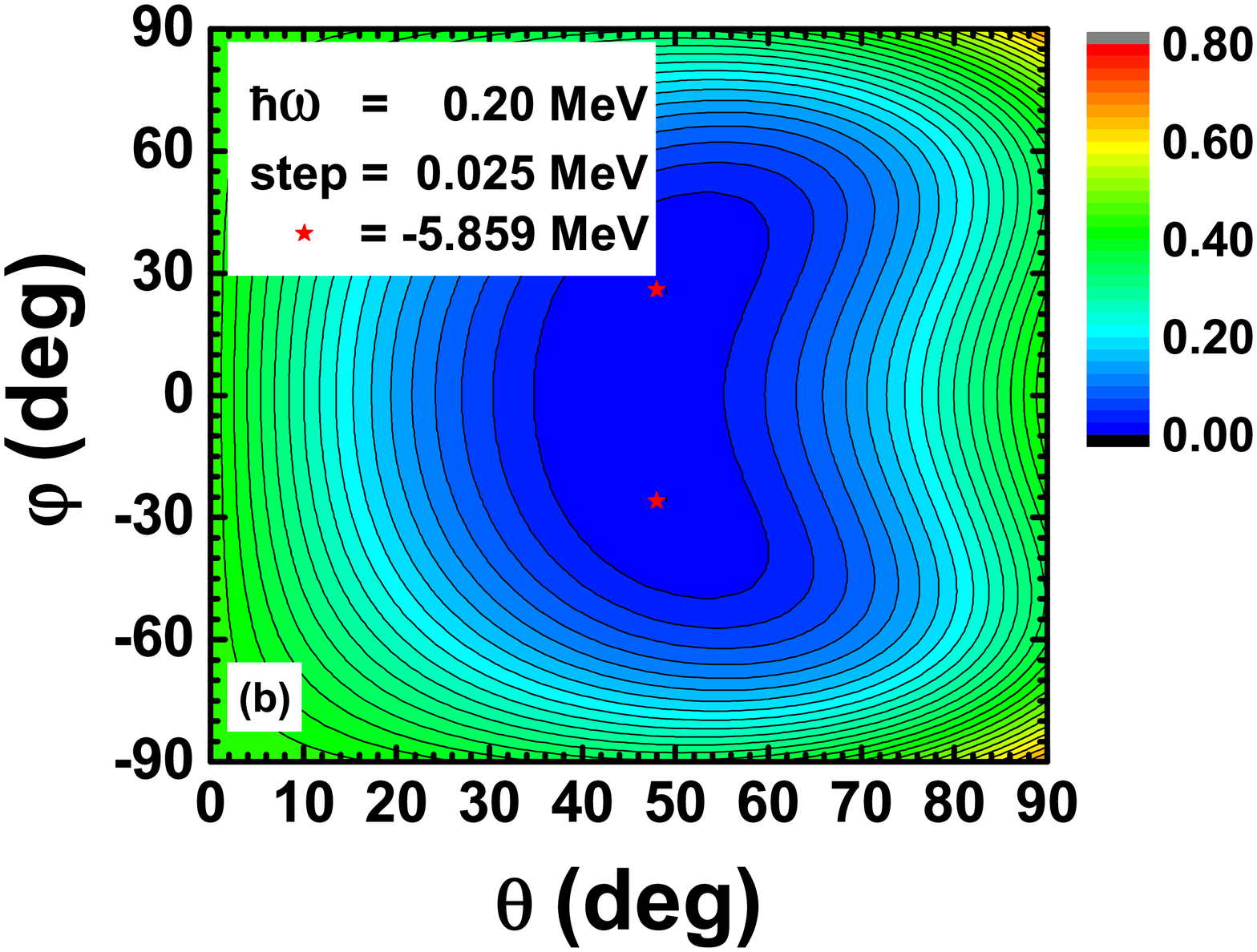}
  \includegraphics[width=7.5 cm]{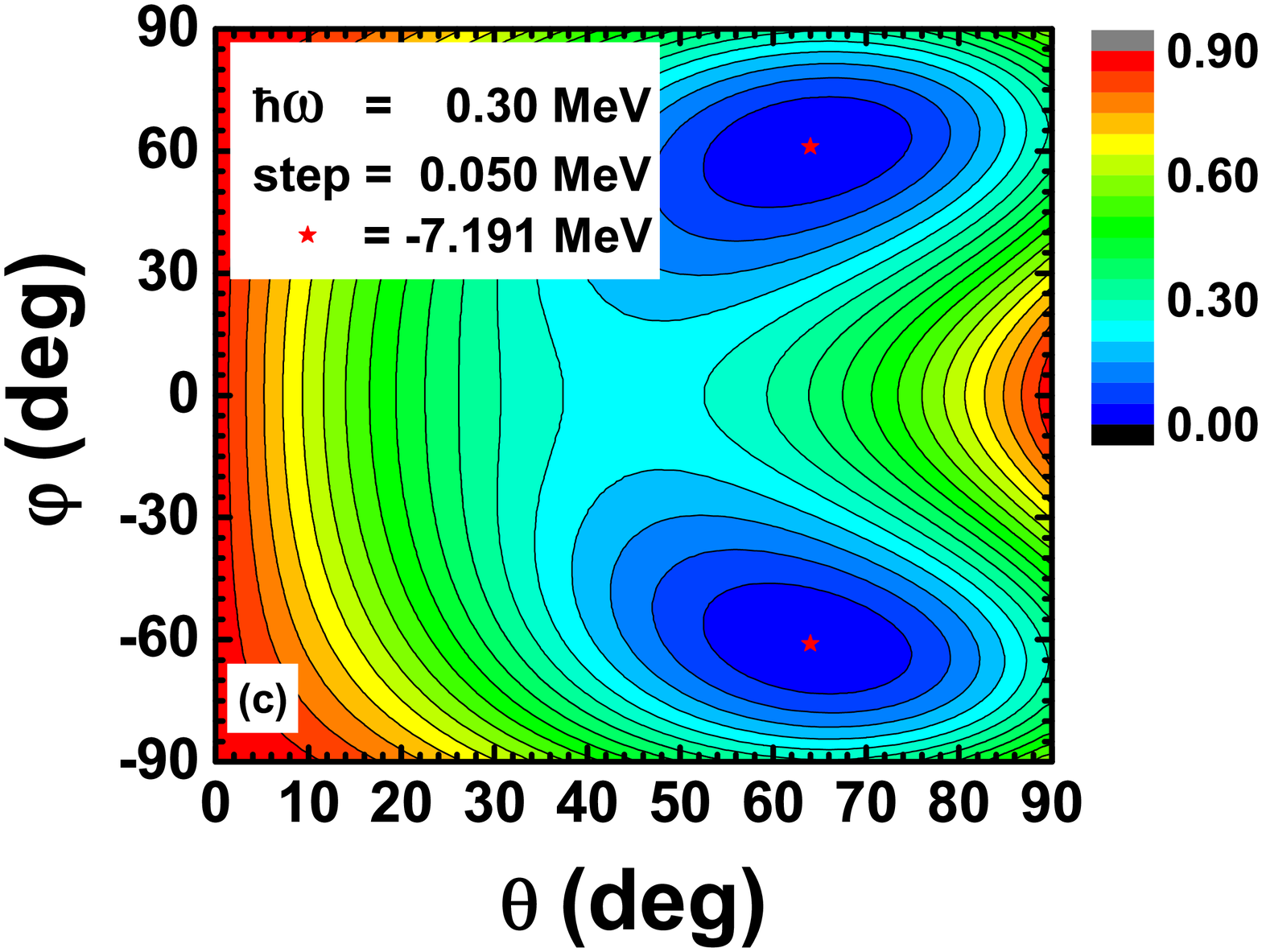}
  \includegraphics[width=7.5 cm]{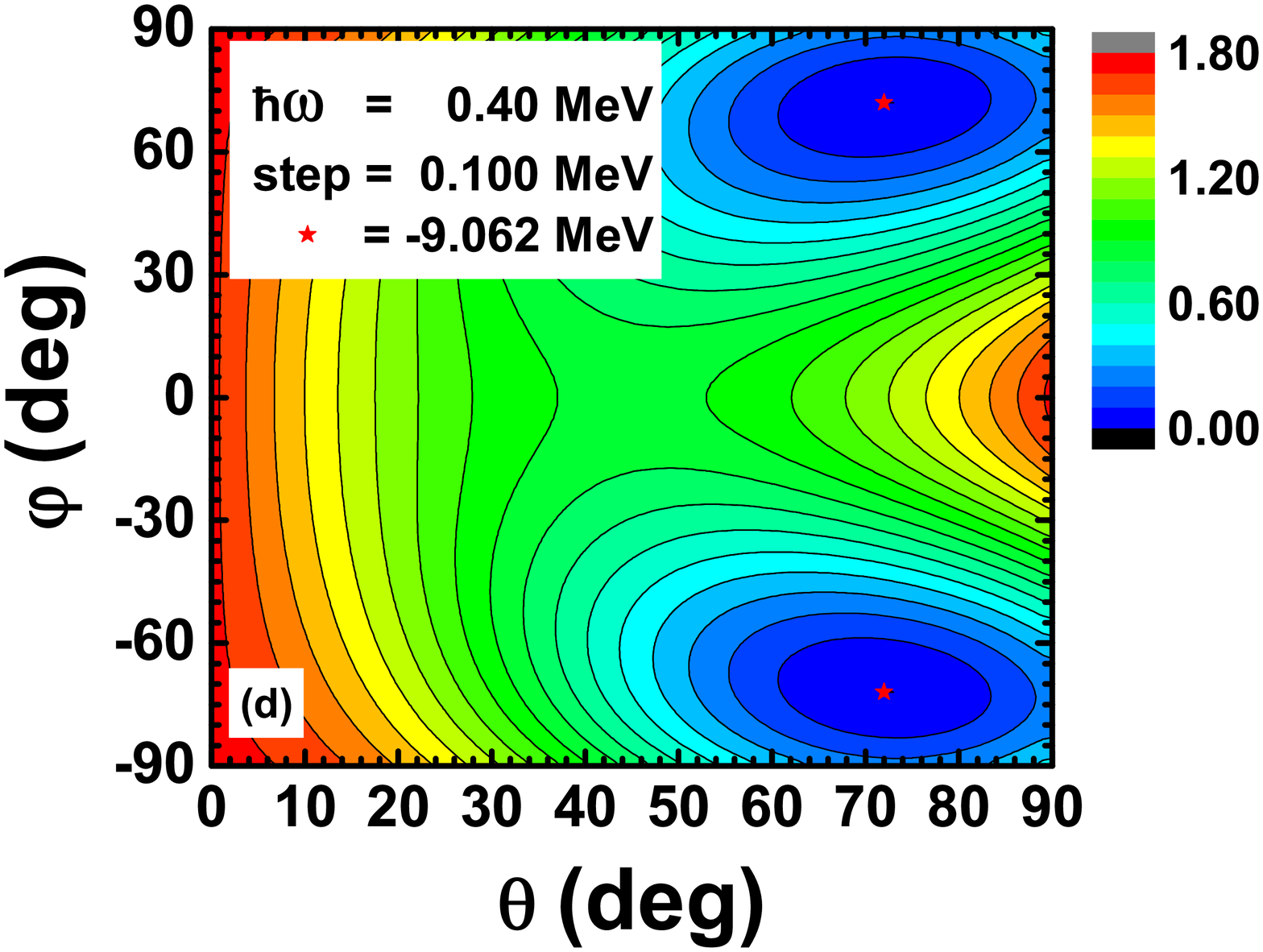}
  \caption{(Color online) Total Routhian surface calculations for the $h_{11/2}$ proton
  particle and the $h_{11/2}$ neutron hole coupled to a triaxial rotor with
  $\gamma=-30^\circ$  at the frequencies $\hbar\omega=0.1$, 0.2, 0.3, 0.4 MeV.
  All energies are normalized with respect to the absolute minimum (star).
  The step is the energy difference between the contour
  lines.}\label{fig1-4}
  \end{center}
\end{figure}

The potential energy surface in the rotating frame, i.e., the total
Routhian $E^{\prime}(\theta,\varphi)$ as a function of $\theta$ and
$\varphi$, is shown in Fig. \ref{fig1-4} at the frequencies
$\hbar\omega=0.1$, 0.2, 0.3, 0.4 MeV. The present results are
consistent with those in Ref.~\cite{Frauendorf1997NPA} where the
total Routhians surface calculations at the frequencies $\hbar\omega=0.10, 0.30, 0.50~\rm
MeV$ have been presented.

One can see that all the potential energy surfaces are symmetrical
with the $\varphi=0^\circ$ line. This means that the two chiral
configurations with $\pm|\varphi|$ for a given tilted angle $\theta$
are identical. With the increasing frequency, the minima in the
potential energy surfaces change from $\varphi=0^\circ$ to
$\varphi\neq 0^\circ$. As discussed in
Ref.~\cite{Frauendorf1997NPA}, this implies the rotating mode
changes from planar to aplanar rotation.

\begin{figure*}[!th]
\begin{center}
  \includegraphics[width=14 cm]{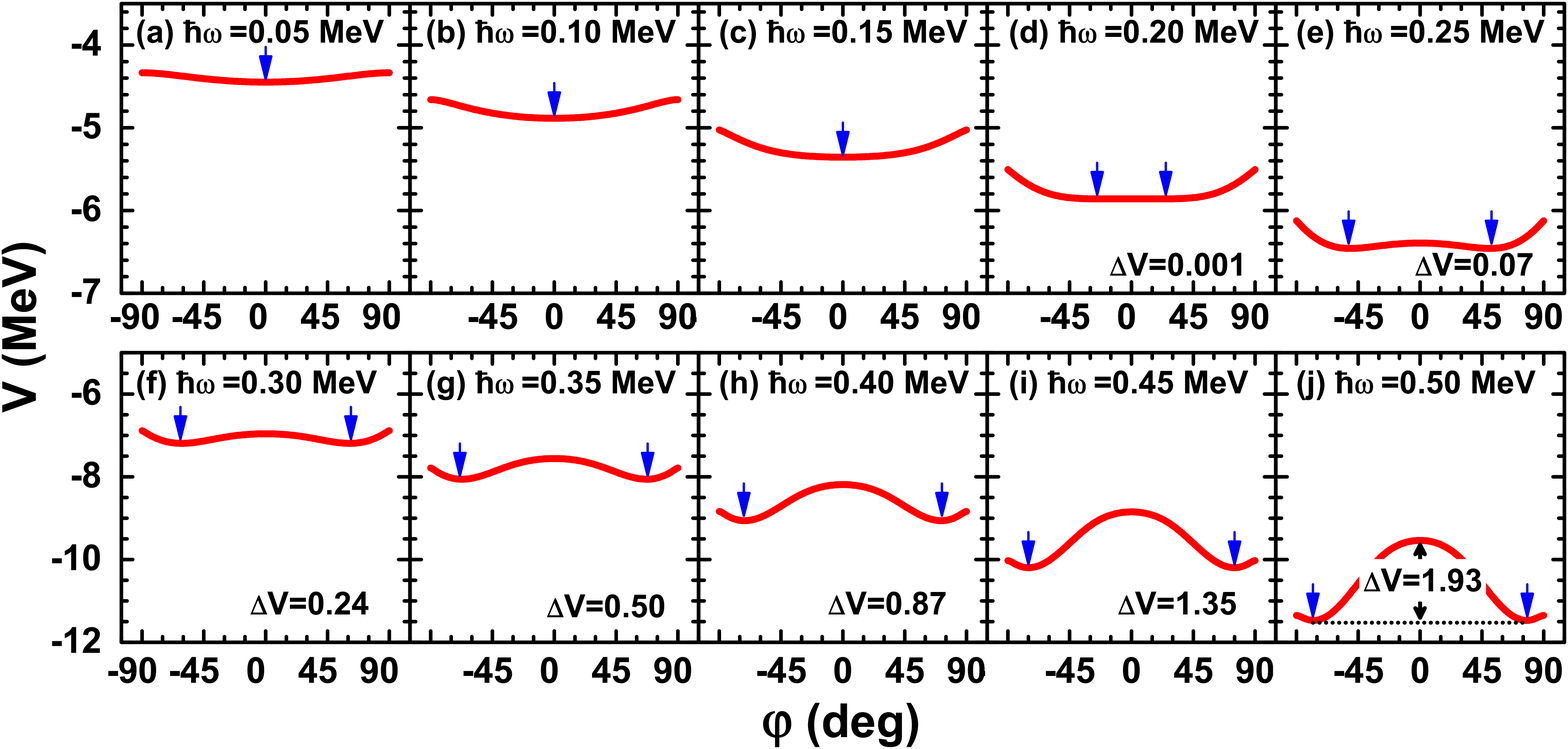}
  \caption{(Color online) The potential energy $V(\varphi)$ as a function of $\varphi$ extracted
  from the total Routhian surface calculations. The arrow labels the position of the potential
  minimum $V_{\rm min}$. The potential barriers defined as $\Delta V=V(0)-V_{\rm min}$ (in MeV)
  are also presented.}\label{fig5}
\end{center}
\end{figure*}

By minimizing the total Routhian $E^\prime(\theta,\varphi)$ with
$\theta$ for given $\varphi$, the potential energy $V(\varphi)$ in
the collective Hamiltonian (\ref{eq13}) is obtained and shown in
Fig.~\ref{fig5}. Again, the potential energy is symmetrical about
$\varphi=0^\circ$ in correspondence with the results shown in
Fig.~\ref{fig1-4}. For the frequency $\hbar\omega \leq 0.15~\rm
MeV$, the potential $V(\varphi)$ is a harmonic oscillator type which
has only one minimum at $\varphi=0^\circ$. This corresponds to the
planar rotation~\cite{Frauendorf1997NPA}. For the frequency
$\hbar\omega\geq 0.20~\rm MeV$, the potential $V(\varphi)$ has two
symmetrical minima. This corresponds to the aplanar
rotation~\cite{Frauendorf1997NPA}. Due to the appearance of the
potential barrier which breaks the chiral symmetry, the stable
chiral solutions are found in the body-fixed frame. The heights of
barrier defined as $\Delta V=V(0)-V_{\rm min}$ (in MeV) with $V_{\rm
min}$ being the value of the potential at the minimum presented also
in the figure. It is found that the potential barrier increases with
the cranking frequency, e.g., from 1 keV at $\hbar\omega=0.20~\rm
MeV$ to about $2$ MeV at $\hbar\omega=0.50~\rm MeV$.

\subsection{Mass parameter}

For chiral rotation with $\hbar\omega\geq 0.20~\rm MeV$, the chiral
vibration frequency $\Omega$ in Eq.~(\ref{eq16}) is taken as
$\Omega=0$, which results in the mass parameter
\begin{equation}
  B(\varphi) = 2\hbar^2\displaystyle \sum_{l\neq 0}\frac{|\frac{\partial \vec{\omega}}{\partial \varphi}
                \langle l|\hat{\vec{j}}|0\rangle|^2}{(E_l-E_0)^3}.
\end{equation}

\begin{figure}[!th]
\begin{center}
  \includegraphics[width=9 cm]{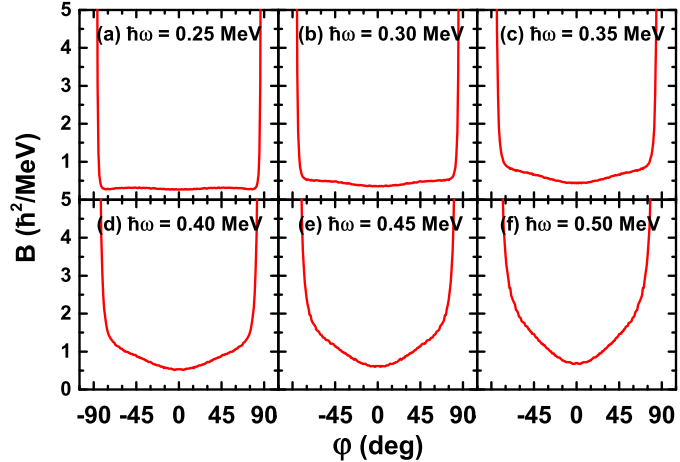}
  \caption{(Color online) The mass parameter $B(\varphi)$ as a function of $\varphi$ for the chiral rotation cases obtained based on TAC.}\label{fig9}
\end{center}
\end{figure}

In Fig. \ref{fig9}, the mass parameter $B(\varphi)$ as a function of
$\varphi$ for the chiral rotation cases is shown. It is seen that
$B(\varphi)$ is symmetric with respect to $\varphi=0^\circ$ and
increases dramatically when $\varphi$ approaches to $\pm 90^\circ$.
In the interior part, $B(\varphi)$ is increased remarkably with
$|\varphi|$ for $\hbar\omega\ge0.35~\rm MeV$, while its dependence
on $\varphi$ is weak for $\hbar\omega=0.25$ and $0.30~\rm MeV$.

\subsection{Energy levels}

The collective Hamiltonian can be constructed by Eq. (\ref{eq13})
with the potential energy $V(\varphi)$ and mass parameter
$B(\varphi)$ obtained.  The diagonalization of the collective
Hamiltonian yields the energy levels and wave functions for each
cranking frequency. As $B(\varphi)$ is quite large at $\varphi\sim
\pm 90^\circ$, a wall approximation is adopted. This corresponds to
the choice of basis in Eq. (\ref{eq17}).

\begin{figure*}[!th]
\begin{center}
  \includegraphics[width=12 cm]{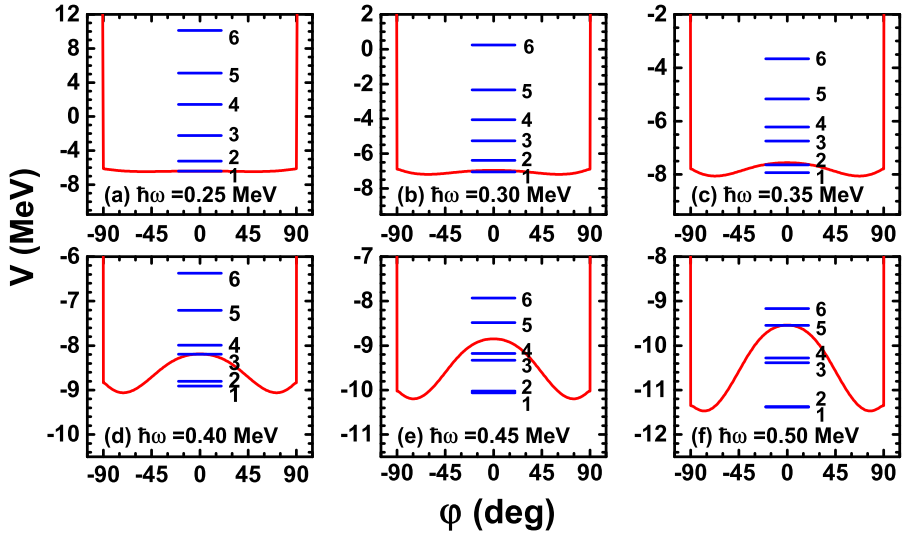}
\caption{(Color online) The six lowest energy levels, labeled as 1-6, obtained from the collective Hamiltonian.
 The potential energy $V(\varphi)$ is included as well.}\label{fig4}
\end{center}
\end{figure*}

In Fig. \ref{fig4}, the six lowest energy levels, labeled as 1-6,
obtained from the collective Hamiltonian are shown together with the
potential energy $V(\varphi)$. It is seen that with the increasing
frequency, the three pairs of energy levels, i.e., levels 1 and 2,
levels 3 and 4 as well as levels 5 and 6, become close to each other.
For example, the energy difference between levels 1 and 2 decreases
from 1.19 MeV at $\hbar\omega=0.25~\rm MeV$ to 0.01 MeV at
$\hbar\omega=0.50~\rm MeV$. Since the potential barrier becomes
higher and wider with the increase of cranking frequency, the
tunneling penetration probability is more and more suppressed.
Therefore, the two levels tend to be more degenerate.

\begin{figure}[!th]
\begin{center}
  \includegraphics[width=8 cm]{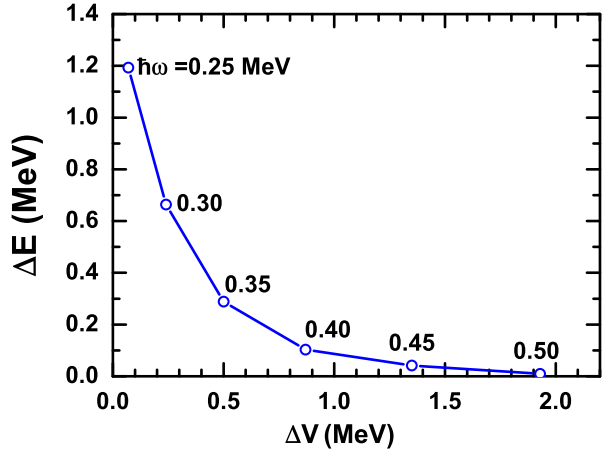}
\caption{(Color online) The energy difference $\Delta E$ between the lowest two levels 1 and 2 of the collective Hamiltonian as a function of the potential barrier height $\Delta V$.}\label{fig11}
\end{center}
\end{figure}

This can be more clearly seen in Fig.~\ref{fig11} where the energy difference $\Delta E$ between the lowest two levels of the collective Hamiltonian is shown as a function of the potential barrier height $\Delta V$. It shows that with the increase of the cranking frequency, the potential barrier grows from 1 keV to about 2 MeV, while the energy differences $\Delta E$ drops correspondingly from 1.19 MeV to 0.01 MeV.

\subsection{Wave function}

\begin{figure}[!th]
\begin{center}
  \includegraphics[width=9 cm]{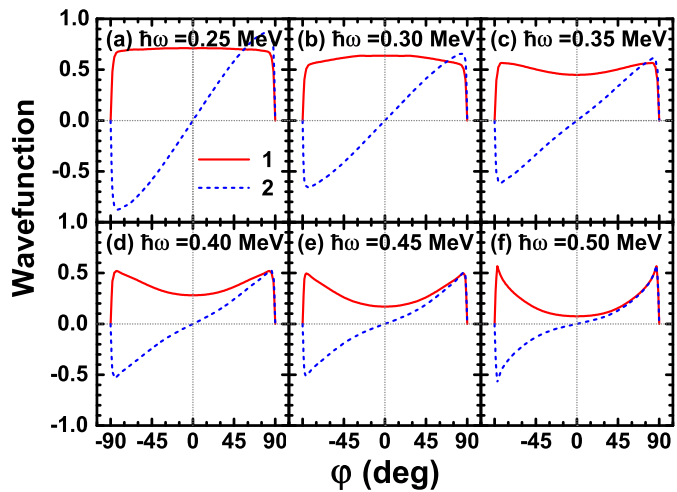}
\caption{(Color online) Wave functions $\psi(\varphi)$ for the lowest two levels 1 and 2 obtained from collective Hamiltonian.}\label{fig12}
\end{center}
\end{figure}

\begin{figure}[!th]
\begin{center}
  \includegraphics[width=9 cm]{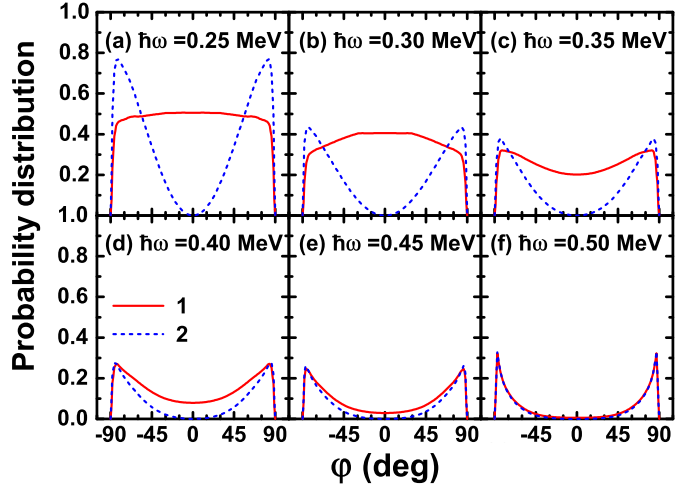}
\caption{(Color online) Probability distributions for the lowest two levels 1 and 2 calculated by $|\psi(\varphi)|^2$.} \label{fig13}
\end{center}
\end{figure}

The wave functions $\psi(\varphi)$ for the lowest two levels are
shown in Fig. \ref{fig12} and the corresponding probability
distributions determined as $|\psi(\varphi)|^2$ are shown in Fig.
\ref{fig13}. It is found that the wave functions are symmetric for
level 1 and antisymmetric for level 2 with respect to $\varphi \to
-\varphi$ transformation. Thus the chiral symmetry broken in the
aplanar TAC solutions is restored. It is also shown that the state
with symmetric wave function is favored in energy compared with the
corresponding state with antisymmetric wave function.

For the frequency $\hbar\omega=0.25$ and $0.30~\rm MeV$, the wave
function of the level 1 remains almost constant except at the border
$\varphi=\pm90^\circ$ which indicates that the fluctuation of the
total angular momentum is large. The wave function of the level 2
shows peaks close to $\varphi=\pm 90^\circ$, which suggests that the
chiral vibration plays an important role for the collective
excitation at the beginning of the chiral rotation region. When the
cranking frequency increases, the wave function of level 1 tends to
show similar pattern. This implies the appearance of the nearly
identical physical properties of these two levels and a good static
chirality appears.

It is well known that the TAC results are given in the body-fixed
frame and the two chiral partner solutions are identical. It cannot
describe tunneling between the left- and right- oriented solutions.
The chiral symmetry is thus broken and only one rotational band can
be obtained. However, in the present formulation of the collective
Hamiltonian, the results are given in the laboratory reference
frame, which could describe the quantum tunneling between the two
chiral solutions. Moreover, the chiral symmetry is restored in the
present framework by considering the contributions from all the
$\varphi$ directions in the potential energy surface.

\subsection{Comparison with exact solutions}

\begin{figure}[!th]
\begin{center}
 \includegraphics[width=9 cm]{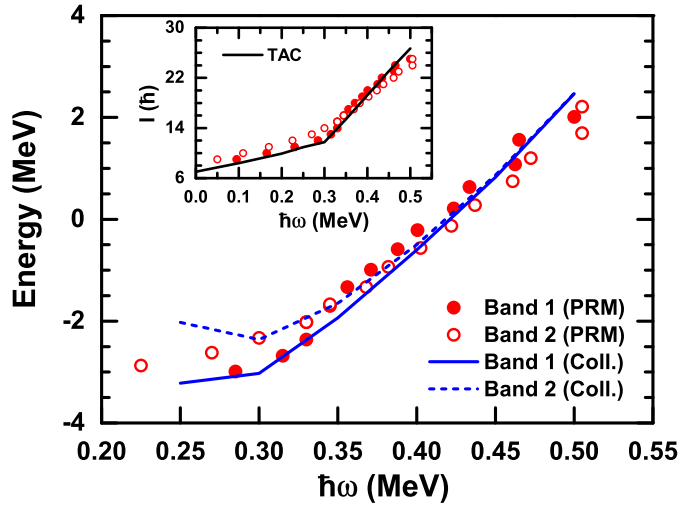}
 \caption{(Color online) The energy spectra of the doublet bands obtained from
        the collective Hamiltonian in comparison with the exact solutions by the PRM.
        Inset: The spin $I(\omega)$ obtained from TAC in comparison with PRM. A similar
        $I(\omega)$ plot has been given in Ref.~\cite{Frauendorf1997NPA}.} \label{fig14}
\end{center}
\end{figure}

The exact solutions for the system discussed here can be
obtained by the particle rotor model. In Ref.~\cite{Frauendorf1997NPA},
the relation between the angular momentum $I$ and the rotational
frequency $\hbar\omega$ as well as the intra band transition
probabilities obtained from the TAC has been compared with the PRM
results. It has been shown that the TAC solutions could reproduce the
results of the yrast levels of the PRM calculation~\cite{Frauendorf1997NPA}.
Here, in the inset of Fig. 8, the spin $I(\omega)$ obtained from TAC in
comparison with PRM is shown once more, where the good agreement can
be clearly seen.

In Fig.~\ref{fig14}, the energy spectra of the doublet bands obtained from the
collective Hamiltonian based on TAC are compared with the PRM results. One
can see that apart from the agreement of collective Hamiltonian and PRM results
for the yrast band, the partner band of PRM can also be reasonably reproduced by
the collective Hamiltonian.

For the collective Hamiltonian results, the energy differences between the doublet
bands become smaller with the increase of the cranking frequency. For the PRM results,
however, the doublet bands become closer up to $\hbar\omega \sim 0.35~\rm MeV$ and the
energy differences between the doublet bands continue to increase for the higher
cranking frequency. As demonstrated both in PRM~\cite{B.Qi2009PLB,B.Qi2009PRC,B.Qi2011PRC}
and TAC+RPA~\cite{Almehed2011PRC} investigations, the doublet bands will attain
a second chiral vibration character, which is not taken into account by the present
collective Hamiltonian investigation.

The present comparison for the energy spectra is made with respect to rotational
frequency $\hbar\omega$, rather than the observable of spin $I$. In the TAC approach,
the angular momentum is not a good quantum number, and the diagonalization of the
collective Hamiltonian is carried out for a given rotational frequency. Therefore,
it is appropriate to compare the energy spectra with respect to the rotational
frequency with the results of PRM. It can be seen from the inset of Fig.~\ref{fig14} that,
for a certain spin, the rotational frequencies of the doublet bands 1 and 2 calculated
by PRM may be quite different (for example, the differences are respectively 0.06, 0.0,
0.02, and 0.04~MeV at $I=12, 16, 20, 24~\hbar$). Here the fluctuation of the
potential energy with $\theta$ is neglected and only the $\varphi$ is treated as
a dynamical variable describing the transition from the left-handed to the right-handed
system. The success of the collective Hamiltonian here guarantees its application for
realistic TAC calculations.


\section{Summary}\label{sec5}

In summary, a collective model which is able to describe the chiral
rotation and vibration is proposed and applied to a system of one
$h_{11/2}$ proton particle and one $h_{11/2}$ neutron hole coupled
to a triaxial rigid rotor. In this framework, it goes beyond the
mean-field approximation, includes the quantum fluctuation in the
chiral degree of freedom, and restores the chiral symmetry. Based on
the tilted axis cranking approach, both the potential energy and
mass parameter as functions of $\varphi$ are obtained and included
in the collective Hamiltonian. Diagonalizing the collective
Hamiltonian with a box boundary condition, the energies and the wave
functions of the collective states corresponding to the motion along
the chiral degree of freedom are obtained. It is found that for
chiral rotation, the partner states become more degenerate with the
increase of the cranking frequency, and for the chiral vibrations,
their important roles for the collective excitation are revealed at
the beginning of the chiral rotation region.


\section*{ACKNOWLEDGMENTS}

This work was partly supported by the Major State 973 Program under
Grant No. 2013CB834400, the National Natural Science Foundation of
China under Grants No. 10975007, 10975008, 11105005, and No.
11175002, the Research Fund for the Doctoral Program of Higher
Education under Grant No. 20110001110087, and the China Postdoctoral
Science Foundation under Grant No. 2012M520101.


\end{CJK*}
\end{document}